\documentclass[english,twocolumn,showpacs,preprintnumbers,amsmath,amssymb,showkeys]{revtex4-2}
\usepackage{graphicx}
\usepackage{dcolumn}
\usepackage{bm}
\usepackage{bigints}
\usepackage{babel,amssymb,amsmath,graphicx,hyperref}

\begin{document}

\title{Uniform skyglow (CMB) observed through the throat of a wormhole}

\author{M.A. Bugaev}
\affiliation{Moscow Institute of Physics and Technology, 9, Institutsky sidestr., Dolgoprudnyi, Moscow region, 141700, Russia}
\author{I.D. Novikov}
\affiliation{Astro-Space Center of P.N. Lebedev Physical Institute, Profsoyusnaya 84/32, Moscow, Russia 117997,}
\affiliation{The Niels Bohr International Academy, The Niels Bohr Institute, Blegdamsvej 17, DK-2100, Copenhagen, Denmark}
\affiliation{National Research Center Kurchatov Institute, 1, Akademika Kurchatova pl., Moscow,  Russia 123182}
\author{S.V. Repin}
\affiliation{Astro-Space Center of P.N. Lebedev Physical Institute, Profsoyusnaya 84/32, Moscow, Russia 117997}
\author{P.S. Samorodskaya}
\affiliation{Moscow Institute of Physics and Technology, 9, Institutsky sidestr., Dolgoprudnyi, Moscow region, 141700, Russia}

\begin{abstract}
        The problem of the possibility of observing a uniform sky glow through the throat of a Morris--Thorne wormhole by
an observer located in another asymptotically flat space-time is considered. It is shown that an individual star has multiple
images, and the image of a luminous sky has a complex structure and contains ring structures. The reasons for
the emergence of such structures are considered. The distribution of radiation intensity in the image along the radial 
coordinate is constructed.In addition, an image of the Morris-Thorne wormhole was constructed against the background of 
uniform sky radiation in the observer's space. A comparison to observations can be made by producing a synthetic 
Morris-Thorne type wormhole image against a CMB background. This image has been construced by a combination of 
the images for inner and outer areas.
\end{abstract}


\maketitle

\section{Introduction}

       Wormholes in the simplest case are one of the static solutions of Einstein's equations. These exotic objects 
consist of two asymptotically flat space-times connected by a throat that lies outside of our three-dimensional space. 
This type of solution was first demonstrated in the paper of A. Einstein \& N. Rosen (Ref.~\citep{Einstein_1935}).  Later, 
many other models of wormholes were proposed both within the framework of general rela\-ti\-vi\-ty  (see 
Refs.~\cite{Ellis_1973, Bronnikov_1973, Morris_1988a, Morris_1988b, Visser_1989, Javed_2022}) and in alternative 
theories of gravity (Refs.~\cite{Agnese_1995, Vacaru_2002, Furey_2005, Eiroa_2008, Botta_2010, DeBenedictis_2012,
Zangeneh_2015, Elizalde_2018, Shaikh_2018, Godani_2020, Singh_2020, Mustafa_2021, Sokoliuk_2022,
Godani_2023}). In recent years, the interest in wormholes and the possibility to observie them has been constantly
growing. The discovery of wormholes would be as great an event in astrophysics as the discovery of the accelerated
expansion of the Universe or the detection of gravitational waves.

       Black holes cannot be observed directly, since these objects do not emit quanta, but only capture them with
their strong gravitational field. Therefore, only the silhouettes of black holes can be observed against the background
of bright objects. The size of this silhouette does not match the size of the event horizon of a black hole, it is much
larger (Refs.~\cite{Repin_2018, Bugaev_2022b, Bugaev_2022c, Malinovsky_2022}). The situation with wormholes is 
radically different. Material bodies can
freely pass through the throat of a wormhole and fall into another asymptotically flat space. It can also be light quanta.
This means that the observer can see the objects of another space through the throat of a wormhole. In this case, 
the silhouette of a wormhole will be significantly different from the silhouette of a black hole. Inside the silhouette of 
a wormhole, characteristic details should be visible that can be interpreted as an image of objects from another space. It 
could be individual stars, galaxies, an accretion disk, or even the cosmological microwave background (CMB). The analysis 
of the image structure should make it possible to distinguish a black hole from a wormhole.

       In this paper, we consider an image of a uniformly luminous sky seen through the mouth of a Morris-Thorne wormhole. 
In addition, we build an image of the same wormhole against the background of a uniformly luminous sky in the space of 
the observer, i.e. an image formed by the rays that do not pass through a wormhole. Finally, we show the resulting 
composite image.

\section{Statement of the problem and geometric parameters of the modell}

         So, our task is to build the image of the luminous sky when observing it through the throat of a wormhole. To do this,
we consider the Morris--Thorne wormhole, whose metric is written as
\begin{equation}
          ds^2 = c^2 dt^2 - dR^2 - \left(R^2 + q^2\right) \left(d\theta^2 + \sin^2\theta\,d\varphi^2\right)\,.
          \label{MT_metric}
\end{equation}
or
\begin{equation}
     ds^2 = dt^2 - \cfrac{r^2}{r^2 - q^2}\,\, dr^2 - r^2
            \left(
               d\vartheta^2 + \sin^2\vartheta \, d\varphi^2
            \right)\,,
          \label{MT_metric2}
\end{equation}
where $c$ is the speed of light, $q$ is a constant.  In equation (\ref{MT_metric2}), c=1 is set and the radial coordinate $r$  
is chosen so that the circumference is equal to~$2\pi r$. When using the $r$ coordinate in equation (\ref{MT_metric2}), 
the trajectories of quanta look more natural and clear.

\begin{figure}[!htb]
  \centerline{
  \includegraphics[width=\columnwidth]{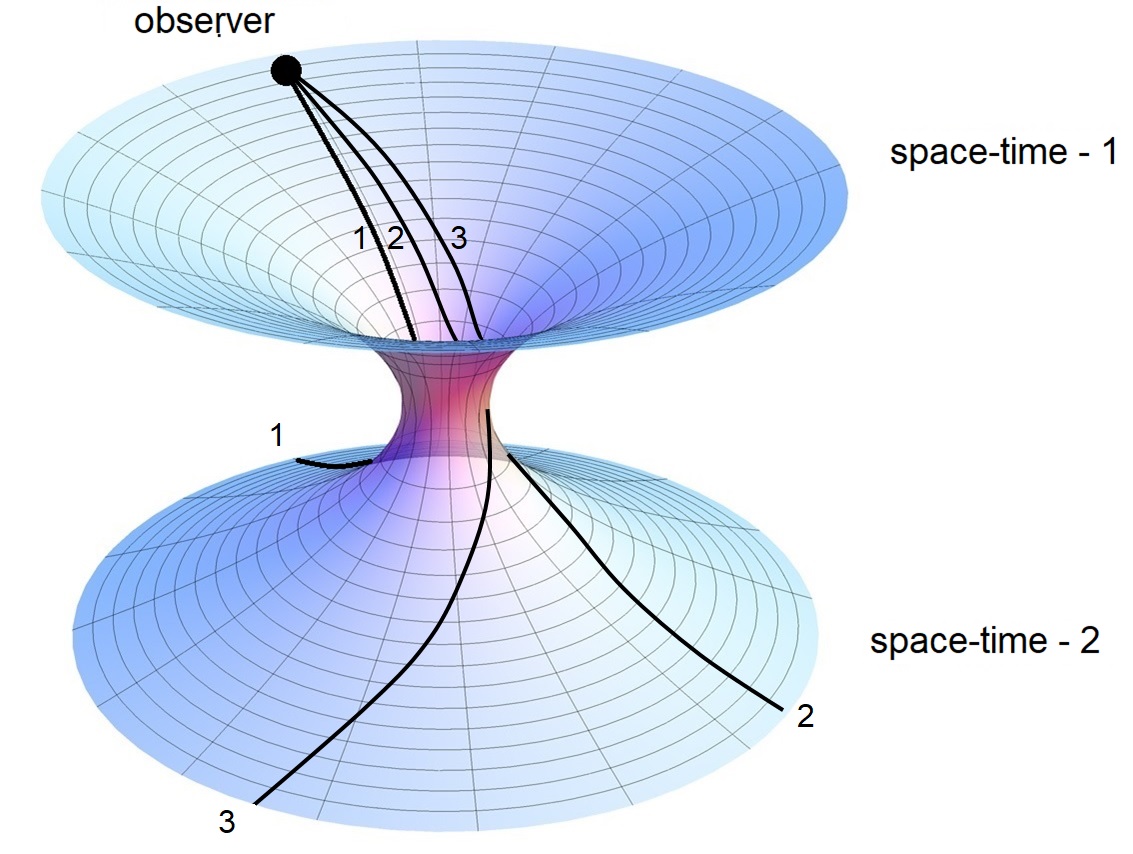}
             }
  \caption{Two asymptotically flat spaces connected by a wormhole. The observer is located in space-1, and
                the luminous sky is located in space-2. Shown here are three null geodesics with different impact parameters
                passing from space-1 to space-2.}
  \label{WH_image_1}
\end{figure}

       Schematically, the location of the observer and the wormhole are shown in~Fig.~\ref{WH_image_1}. We will 
conditionally say that the observer is in space-1, far from the wormhole (at "infinity"{}), and the radiation comes to him from
the celestial sphere, located in space-2, through the throat of the wormhole. In addition to the observer and the celestial
sphere, Fig.~\ref{WH_image_1} qualitatively shows the mutual arrangement of the trajectories of three photons that come
to the observer from space-2. The quanta trajectories are bent due to passing through the throat of the wormhole. With 
the path of the beam following higher curvature as it’s impact parameter approaches the boundary of the wormhole throat. Further in 
the paper, we will discuss the mutual arrangement of the trajectories and the degree of their curvature in more detail.

         To construct an image, it is more convenient to assume that light does not propagate from the object (celestial sphere) to
the observer, but, on the contrary, from the observer to the object, while passing through the throat of the wormhole.
This method of calculation is justified, since in the metric (\ref{MT_metric2}) the beam reversibility principle is fulfilled.

\section{Equations of motion and numerical methods }

       To construct an image, it is necessary to construct a large number of quantum trajectories (zero geodesics) in
the Morris--Thorne metric. Geodesic equations can be obtained by separating the variables in the Hamilton-Jacobi equation:
\begin{equation}
  \begin{split}
               \left(\cfrac{\partial S}{\partial t} \right)^2  -
                \cfrac{r^2 - q^2}{r^2} \left(\cfrac{\partial S}{\partial r} \right)^2 -
                \cfrac{1}{r^2} \left(\cfrac{\partial S}{\partial \theta} \right)^2 -  \\
              - \, \cfrac{1}{r^2 \sin^2\theta} \left(\cfrac{\partial S}{\partial \varphi} \right)^2 - m = 0 ~ ~ ~\,,
  \end{split}
\end{equation}
where $S$ is the action, $m$ is the mass of the test particle. We are looking for the solution of this equation in the form:
\begin{equation}
               S = -Et + L\varphi + S_\theta (\theta) + S_\varphi(\varphi),
\end{equation}
where the constants $E$ and $L$ denote the energy and angular momentum of the test particle at infinity.

        Trajectories of quanta are obtained by numerical integration of the geodesic equations in the Morris--Thorne
metric. These equations are derived by separation the variables in the Hamilton--Jacobi equation and in dimensionless 
variables: $\tilde r = r/q$, $\tilde t = t/q$, $\tilde E = E/q$, $\tilde L = L/mq$, $\tilde Q = Q/m^2q^2$ have 
the form (tildes are omitted for brevity) (Refs.~\cite{Zakharov_1994, Zakh_Rep_1999})):
\begin{eqnarray}
      \cfrac{dt}{d\sigma} & = & \cfrac{1}{r^2}\,\,, \label{Eq_motion2_1}  \\
      \cfrac{dr}{d\sigma} & = & r_1\,, \label{Eq_motion2_2}  \\
      \cfrac{dr_1}{d\sigma} & = & 2 \left(\eta - \xi^2\right) r^3 - \left(1 + \eta + \xi^2\right) r\,, 
                                \label{Eq_motion2_3} \\
     \cfrac{d\theta}{d\sigma} & = & \theta_1\,, \label{Eq_motion2_4} \\ 
     \cfrac{d\theta_1}{d\sigma} & = & \cfrac{\xi^2\cos\theta}{\sin^3\theta}\,\,,\label{Eq_motion2_5} \\
      \cfrac{d\varphi}{d\sigma} & = & \cfrac{\xi}{\sin^2\theta}\,\,.  \label{Eq_motion2_6}
\end{eqnarray}
where $t,1/r,\theta,\phi$ are the Boyer-Lndqvist coordinates, $\eta = Q/M^2E^2$ and $\xi = L_z/ME$~ are
the Chandrasekhar constants. $Q$~-- Carter separation constant (Ref.~\cite{Carter_1968}).  Using the radial 
coordinate~$1/r$ instead of the traditional~$r$ is more convenient for numerical integration, since a significant part of 
the quantum trajectory lies at a large distance from the wormhole.

\section{Uniform sky glow seen through a wormhole.  Stargazing through the throat of a wormhole}

       Let us consider a uniform glow in space-2 across the whole sky.  An example of such a glow can be microwave 
background radiation in space-2, which comes to a given point from the solid angle $4\pi$. One can also imagine a starry 
sky in which the stars are densely packed so that the disks of the individual stars are indistinguishable.  Our goal will be 
to construct an image of this sky as it is seen by an observer in space-1 through the throat of a wormhole.

\begin{figure*}[tbh]
  \centerline{
  \includegraphics[width=8cm]{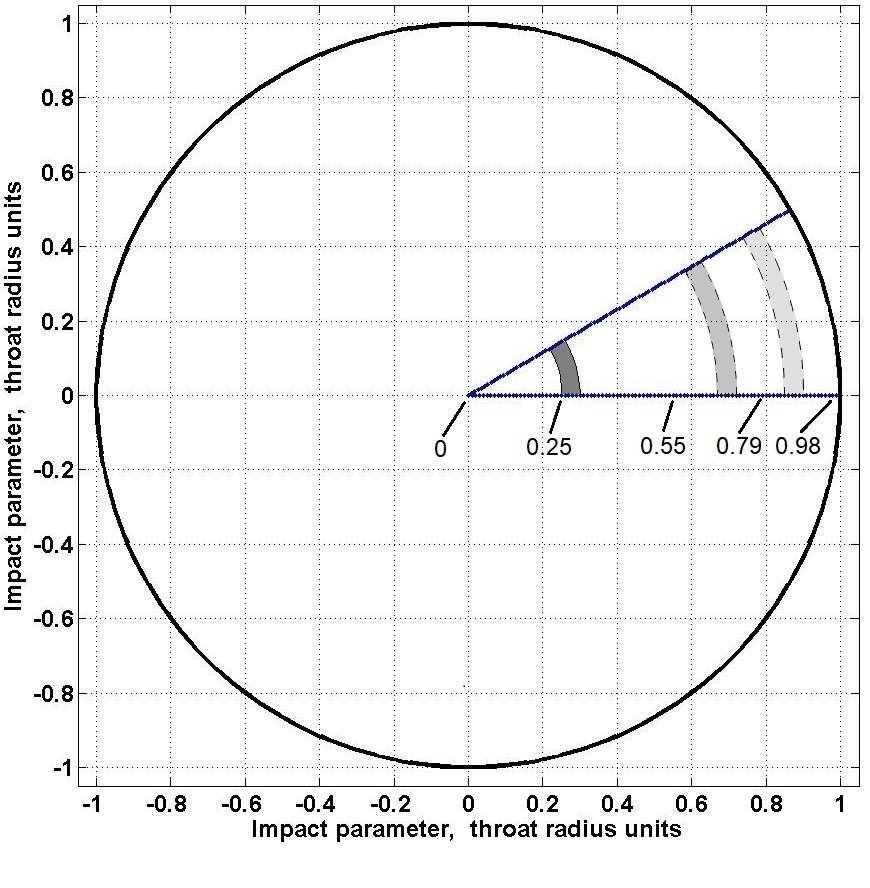}
  \includegraphics[width=8cm]{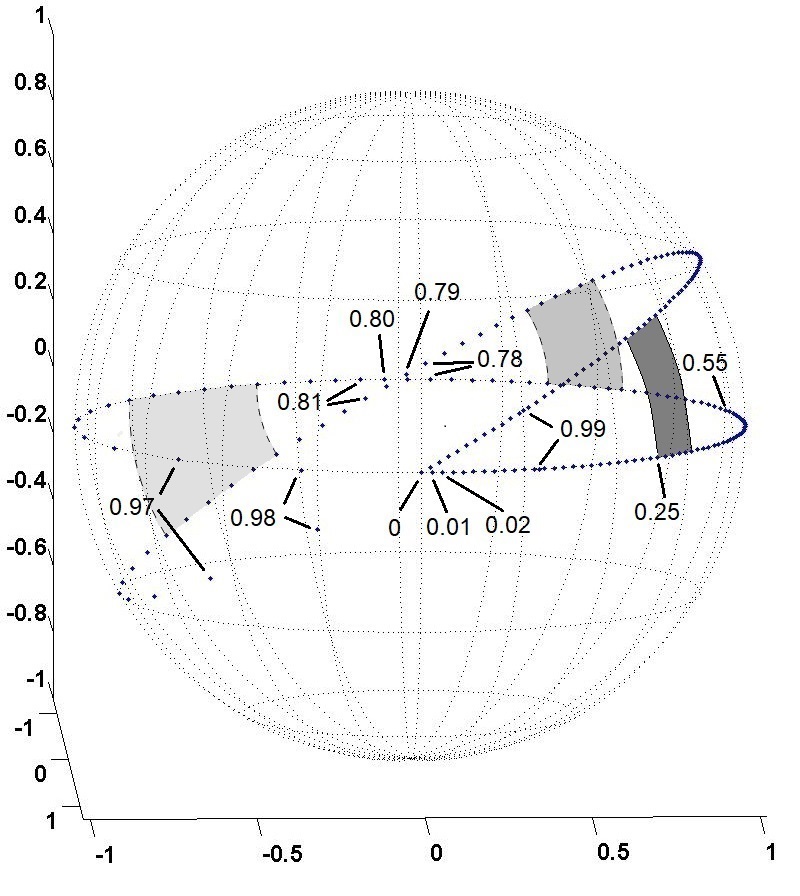}
             }
  \caption{Deviation of quanta with different impact parameters when passing through the throat of a massless wormhole.
                 The circle in the left panel is an image of the throat of a wormhole.    See text for details.}
  \label{WH_impact_param}
\end{figure*}

       Let us consider first the general ideas of image formation. As we noted above, it is more convenient to consider
the trajectories of quanta that propagate from the observer to the source.  Let 100~quanta be sent from the observer 
in space-1 to the wormhole, which are "targeted"{} at 100 different points in the image of the throat, lying along one 
radius of the image. Thus, let the quanta impact parameters increase uniformly from 0 to $0.99q$, where $q$~ is 
the throat radius. In polar coordinates in the image of the throat all these points fall on the radius with $\varphi=0$.

       One hundred quanta sent along the radius, having passed through the throat of the wormhole, should fall in space-2 on
the arc of the great circle (a straight line on the sphere). However, the angular distances between these quanta will no longer
be constant, since the closer the quantum's impact parameter is to the throat radius, the more its trajectory is bent. In 
Fig.~\ref{WH_impact_param} on the left, dots show the quanta along the radius with their impact parameters $b$, and on 
the right, the position of the same quanta in the equatorial plane on the celestial sphere of space-2, where the corresponding 
impact parameters $b$ are also indicated.

       Now let's send another 100 quanta from the observer towards the wormhole, but located along a different radius, inclined to
the original at some angle, for example, $\alpha=30$ degrees.  Such a radius is shown in Fig.~\ref{WH_impact_param} in the
left panel. On the celestial sphere in space-2, these quanta will also fall on the arc of the great circle, inclined to the original at
the same angle. The location of these quanta is shown by dots in the right panel in Fig.~\ref{WH_impact_param}.  Two arcs of 
a great circle, as it is known, intersect at two diametrically opposite points of the celestial sphere.

       After such reasoning, it becomes clear which parts of the celestial sphere of space-2 we see inside the sector shown in 
Fig.~\ref{WH_impact_param} on the left. A fragment of a narrow ring inside the silhouette of a wormhole is formed by quanta 
emitted in space-2 from the corresponding marked portion of the celestial sphere located between two large circles. Three pairs of 
such fragments corresponding to each other are shown with different shading on both panels in Fig.~\ref{WH_impact_param}.

       The area of a segment of a narrow ring bounded by two radii and two arcs of a circle in the throat of a wormhole is equal to
$$
              dS = \alpha b\, db,
$$
where $b$ is the value of the impact parameter ($b < 1$), $db$ is the width of the ring, and $\alpha$ is the angle between
the radii. As the impact parameter $b$ increases, this area increases linearly.

        The area of the corresponding portion on the celestial sphere ("quadrilateral"{}) can be approximated with good accuracy 
as the sum of the areas of two spherical triangles with known vertex coordinates. This area first increases (linearly at the very 
beginning) with 
the growth of $b$, then reaches a maximum and begins to decrease. On the opposite side of the celestial sphere, this area 
decreases to zero and begins to increase again. Such a "periodic process"{} can continue for an arbitrarily long time, but 
the distances between the points of arrival of quanta will increase monotonically. It is clear from Fig.~\ref{WH_impact_param} 
that the area reaches its maximum value at $b\approx 0.45$, its minimum value between $b=0.79$ and $b=0.80$, another 
maximum at $b\approx 0.93 \div 0.94$ and, finally, the second low approximately in the middle between $b=0.98$ and $b=0.99$.
If we compare Fig.~\ref{WH_impact_param} with the schematic Fig.~\ref{WH_image_1}, we can see that the trajectory~1
corresponds to the impact parameter $b_1=0$, trajectory~2~ to the impact parameter $b_2 = 0.793$, and trajectory~3~ to 
$b_3\approx 0.93$.

       The brightness of the image, seen through the throat of the wormhole, is proportional to the ratio of the areas of 
the quadrilateral on the celestial sphere and the fragment of the corresponding narrow ring in the throat of the wormhole.
It is clear that near the values of the impact parameter $b=0.793$ and $b=0.985$ the intensity minimum will be observed in 
the image, and near $b=0.93$~-- the maximum. This "rule" {} fails only for $b\approx 0.45$, since in this case both areas start 
increasing from zero.

       With further approaching of the impact parameter to the radius of the wormhole throat, the light quantum can make more 
and more revolutions in the throat. Theoretically, the number of revolutions can increase indefinitely. Therefore, near 
the boundary of the wormhole silhouette, an infinite number of alternating bright and dark rings with ever-decreasing width will 
be observed.

       So, we can qualitatively explain the formation of ring structures in the image of a uniformly luminous sky seen through 
the throat of a wormhole. Similar ring structures should also form on the outer side of the wormhole silhouette. In this case, 
the observer sees an image of the luminous sky of his space.

\begin{figure}[htb]
  \centerline{
  \includegraphics[width=0.95\columnwidth]{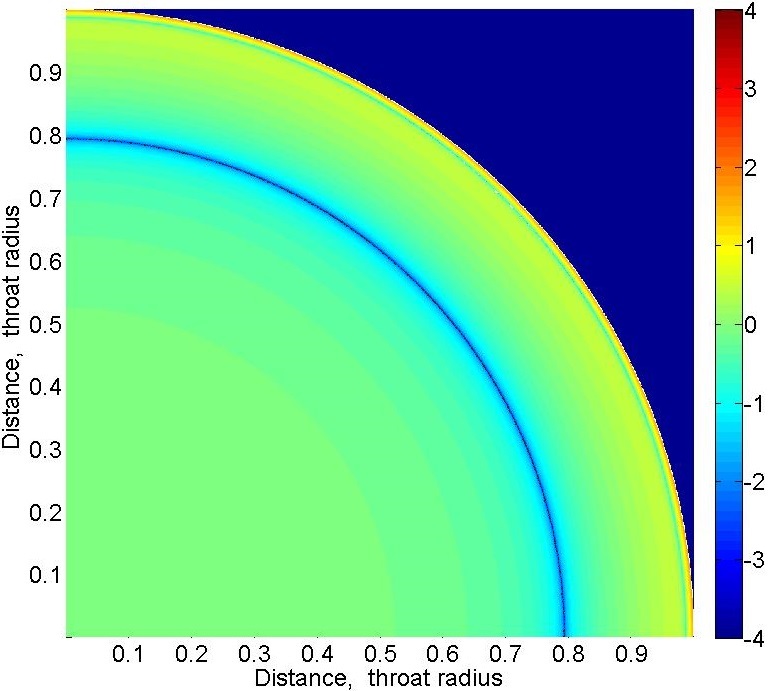}
             }
  \caption{An image of a uniformly luminous sky as viewed through the throat of the Morris-Thorne wormhole. 
                 The intensity of the observed emission is shown on a logarithmic scale.}
  \label{WH_image_3}
\end{figure}

\begin{figure*}[hbt]
  \centerline{
  \includegraphics[width=8.5cm]{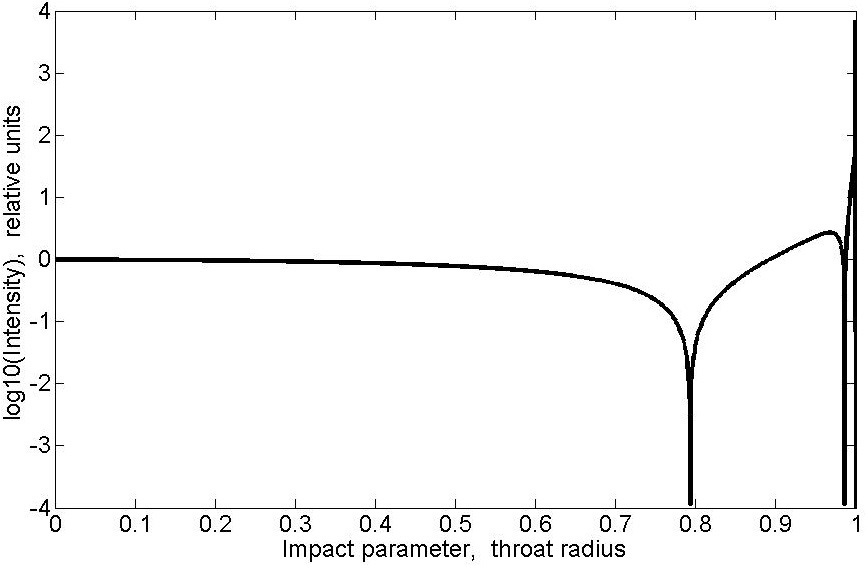} ~ ~ 
  \includegraphics[width=7.5cm]{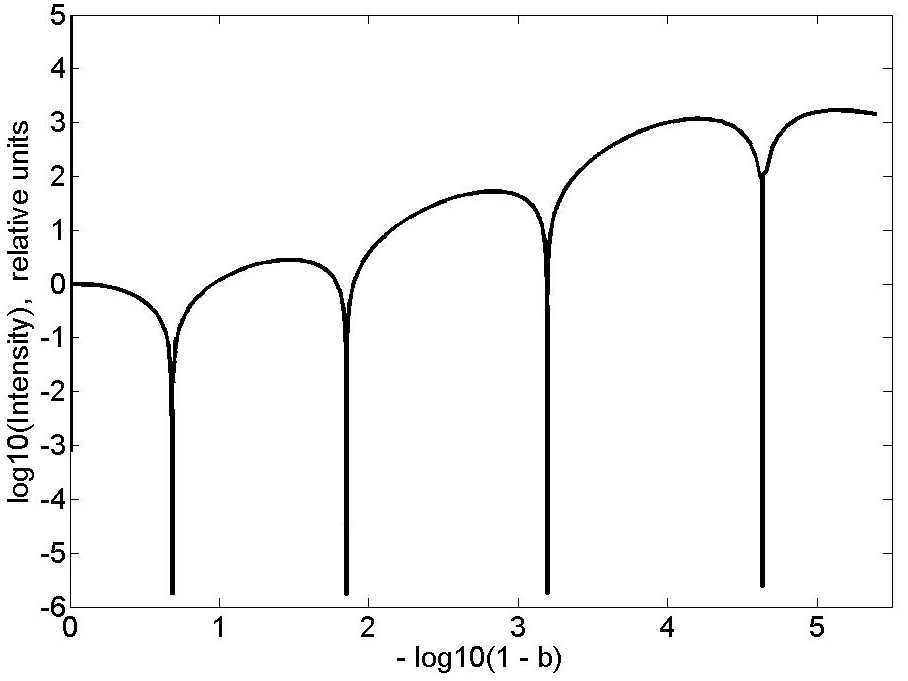}
             }
  \caption{Plot of emission intensity along the radial coordinate. 
                 The intensity distribution corresponding to the image in Fig.~\ref{WH_image_3} is shown on the left. 
                 The same distribution is shown on the right, but the value $-\log_{10}(1-b)$ is used as the radial 
                 coordinate, where $b$~ is the impact parameter.}
  \label{WH_intense_plot_inner}
\end{figure*}

        Fig.~\ref{WH_image_3} shows the image of a uniformly glowing sky seen through the throat of a wormhole. This image 
has central symmetry, so it is enough to show one-fourth of it. False colors in the palette are proportional to the logarithm of
brightness, and the color bar explains them.   The image clearly shows a low-brightness ring with a radius of 0.793 of 
the throat radius. Then 
the brightness increases and near the very edge of the silhouette again passes through a minimum.  The plot of the image 
intensity along the radial coordinate is shown in Fig.~\ref{WH_intense_plot_inner} on the left. Theoretically, the number of 
dark and light rings should be infinite, and their width should decrease. It is impossible to show more than two rings in 
the figure, since their width becomes less than a pixel size. A more complete representation, however, can be shown if 
the value $\log_{10}(1-b)$ is used as the radial coordinate, where $b$~ is the impact parameter. The intensity distribution 
using such a radial coordinate is shown in Fig.~\ref{WH_intense_plot_inner} on the right. Four minima are clearly visible on 
the graph, and, moreover, it is clear that when approaching the edge of the image, the intensity increases on average.
We emphasize that the observer measures the intensity distribution along the wormhole silhouette radius shown in 
Fig.~\ref{WH_intense_plot_inner} on the left. The graph shown in Fig.~\ref{WH_intense_plot_inner} on the right helps 
to understand the details of the structure of this distribution.

        In order to show more rings in the image, it is necessary to significantly increase the relative accuracy of integrating 
the equations of motion or immediately write these equations using the specified coordinate. But the general view of 
the image of the luminous sky through the throat of the wormhole is already clear. It contains an infinite number of rings of 
variable intensity, the width of which decreases as the impact parameter approaches the radius of the throat, while 
the brightness of the rings increases.

\begin{figure*}[htb]
  \centerline{
  \includegraphics[width=14cm]{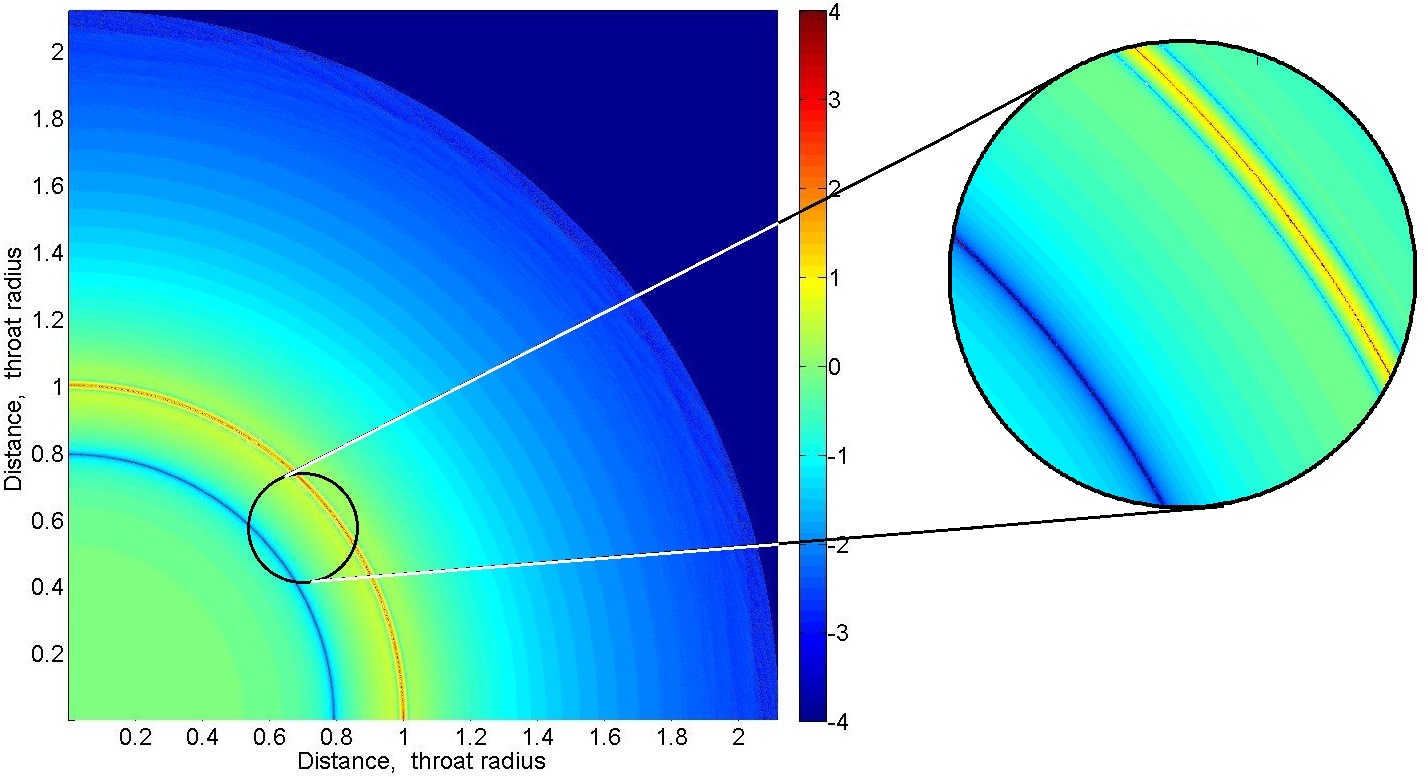}
             }
  \caption{An image of a uniformly luminous sky as viewed from both the inside and 
                 the outside of a Morris--Thorne wormhole. The intensity of the observed 
                 radiation is shown on a logarithmic scale.}
  \label{WH_image_5}
\end{figure*}

{}  {}

\section{Uniform sky glow in both spaces }

        From the outer side of the silhouette of the wormhole, we should also see the ring structures. The reason for their formation 
is exactly the same as when the light passes through the throat of a wormhole. The only difference is that now the observer sees 
the objects located in his space-time.

        The structure of the image of the luminous sky, observed simultaneously from the outer and inner sides of the wormhole, is 
shown in Fig.~\ref{WH_image_5}. In this case  it is also sufficient to show one quarter of the image due to the symmetry. Near 
the boundary of the silhouette of the wormhole, a ring of low intensity can be seen. Theoretically, there should also be an infinite 
number of rings on the outside near the boundary. However, it is impossible to show them at the selected scale, because all these 
rings together, starting from the second one, have a width too small to be visible at such scales.

         The positions of the outer rings with good accuracy turn out to be symmetrical to the inner rings with respect to the wormhole 
boundary. Thus, pairs are formed of rings with impact parameters $b_1 = 0.986$ and $b_2 = 1.0158$, as well as 
$(0.99936, 1.00065)$ and $(0.999978, 1.000026)$. There is no "pair"{} only to the ring at a distance of 0.793 from the center of 
the wormhole.

\begin{figure}[htb]
  \centerline{
    \includegraphics[width=0.9\columnwidth]{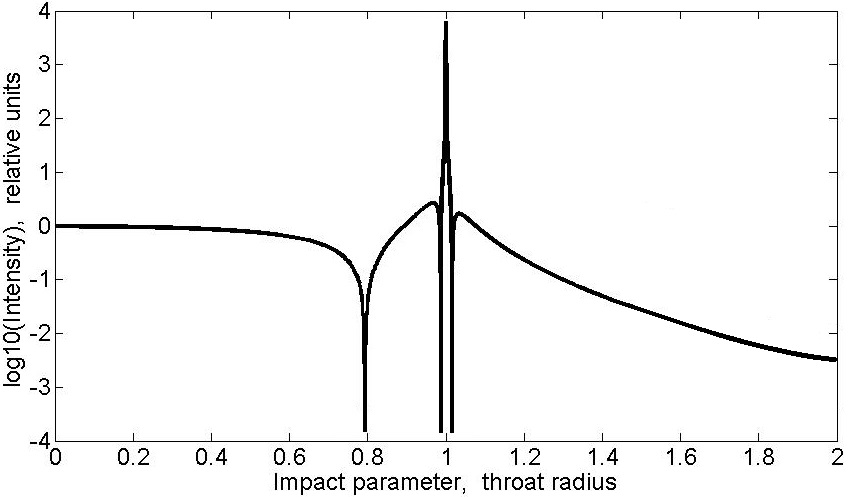} 
             }
  \caption{.}
  \label{WH_image_6}
\end{figure}

          The intensity distribution along the radial coordinate for the inner and outer parts of the wormhole silhouette is shown 
in~Fig.~\ref{WH_image_5}. The graph clearly shows already familiar intensity minima at the distances of 0.793, 0.986 and 1.016 in 
units of the wormhole radius. In addition, a general increase in intensity is clearly visible when approaching the boundary. It is 
impossible to show more minima on such a scale, but the overall picture is completely similar to that shown in 
Fig.~\ref{WH_intense_plot_inner}, i.e. there are intensity minima in the image, their width and the distance between them 
decrease, and the overall brightness of the image increases when approaching the edge of the wormhole.  It would be possible 
to plot the intensity with a logarithmic coordinate along the radius for the outer part of the image as well, but its appearance is almost 
the same as that shown in Fig.~\ref{WH_intense_plot_inner} on the right.

         A wormhole connects two asymptotically flat spaces. The second exit of the wormhole can be located both in our Universe and 
in some other universe connected with our one by such a "bridge"{}. In the latter case, we see the objects from another universe
through the throat of a wormhole. In this universe, the intensity of the microwave background can be very different from what is in
our Universe. And this means that the distribution of radiation intensity from the inner and outer sides of the wormhole silhouette 
may not be related to each other. Figures~\ref{WH_image_5}, \ref{WH_image_6} show the intensity distribution for the case when
both exits of the wormhole are in the space-time of our Universe. It is clear that at a large angular distance from the silhouette of 
the wormhole, the radiation intensity remains unperturbed and equal the intensity of the glow of the sky in our space. From
Fig.~\ref{WH_image_5}, \ref{WH_image_6} it follows that the brightness of the image in the center of the wormhole silhouette is
much higher than the brightness of the surrounding background. This difference is more than two orders of magnitude. If the second
exit of the wormhole lies in another universe, then the intensity in the center of the image can even have an arbitrary value. 
However, in any case, the observer should see an increase in image brightness when approaching the edge of the wormhole 
silhouette and the low-intensity ring at the distance of 0.793 wormhole radius.  The observation of other rings requires very high
angular resolution.

\begin{figure*}[htb]
  \centerline{
    \includegraphics[width=10cm]{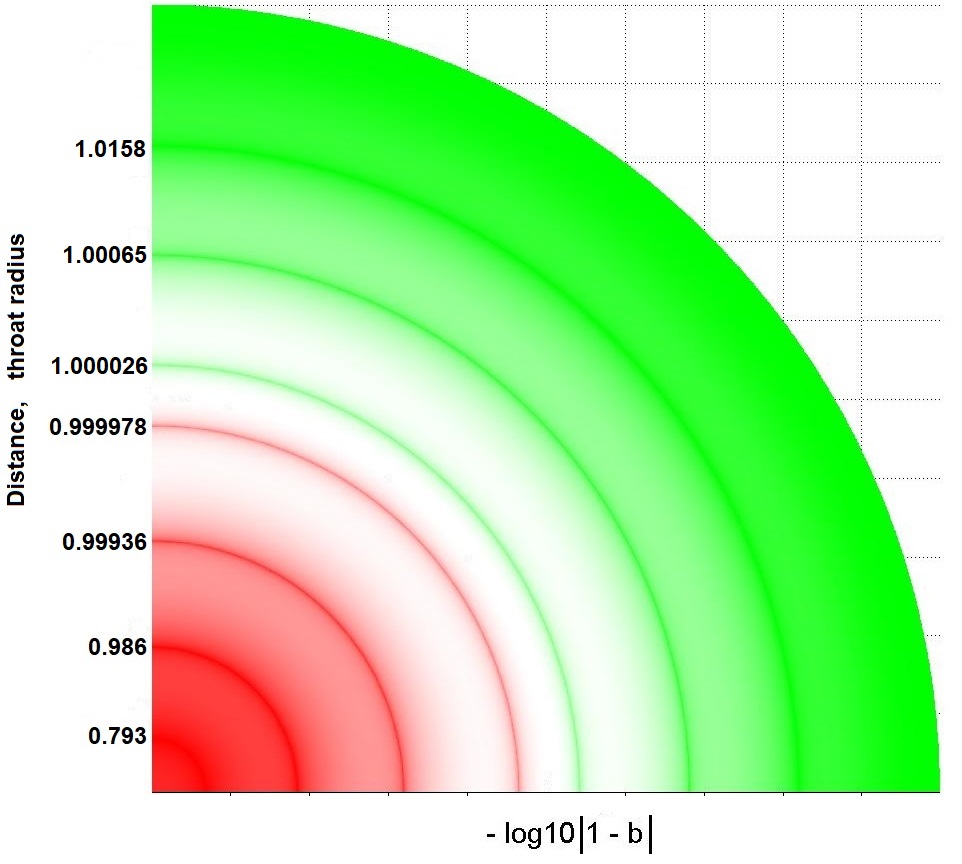} ~
             }
  \caption{An image of a uniformly luminous sky viewed from both the inside and the outside of a Morris-Thorne 
                 wormhole. The value $-\log_{10}|1-b|$ is used as the radial coordinate along both axes. 
                The radii of the dark rings are indicated on the left. }
  \label{WH_image_7}
\end{figure*}

         An even more complete idea of the visible image of a wormhole is provided by Fig.~\ref{WH_image_7}, in which 
$-\log_{10}|1-b|$ is used as the radial coordinate. By using such a coordinate, the resulting ring structures can be shown in detail. 
The image of the inner and outer parts of the wormhole is shown in different palettes to emphasize the fact that 
the brightnesses of these parts can be independent if the second exit of the wormhole that lies in a different universe. The values of 
the radii of the dark rings are indicated in the figure by conventional units. The light areas of the figure correspond to the brighter 
parts of the wormhole image. In Fig.~\ref{WH_image_7} it is clearly seen that the brightness of the image increases when
approaching the edge of the wormhole. In addition, one can notice that the relative increase in the brightness of the rings is
approximately the same for the inner and outer parts of the image.

         The constructed images of a wormhole can be used in observations to distinguish these objects from black holes.

\section{Conclusions}

       Wormholes may possibly exist in the centers of galaxies (Ref.~\cite{Kardashev_2020}).

       In the considered model of the image of a massless wormhole, there are characteristic details that can be used to identify these
objects by interferometric observations (Ref.~\cite{Mikheeva_2020}). Thus, quite complex ring structures with varying image 
brightness can be observed inside the silhouette of a wormhole. In this way, images of wormholes are fundamentally different 
from images of black holes. In the paper we found the radii of dark rings that appear in the image.

       An increase in intensity when approaching the wormhole boundary both from the inside and outside can also serve as a characteristic 
feature of such an object. In other words, a bright ring should be observed near the boundary of a single wormhole. The intensity 
distribution in such a ring along the radial coordinate differs from what can be observed in the case of the outer edge of the black 
hole silhouette, and this information can be used by observations. Perhaps such observations can be organized in future space projects.

        It is also interesting that the brightness of the image of the inner part of the wormhole exceeds the brightness of the background by 
more than two orders of magnitude. This means that a wormhole can be observed not as a "shadow"{}, but as a bright object, if it is passable 
and its second exit is in our Universe. However, if the other exit of the wormhole does not lie in our universe, then the brightness of 
the central part of the image and the presence of dark rings in it can give important information about the physical processes in another 
universe.

        To search for and observe wormholes, one can use the catalog of supermassive black holes 
(Refs.~\cite{Mikheeva_2019, Malinovsky_2022}), in the centers of galaxies. There are many candidates in this catalog for carrying out 
the observations with ground-based interferometers. The discovery of such objects will undoubtedly give a new impetus to 
the develop+ment of astrophysics.

\begin{acknowledgements}
     S.R. expresses his gratitude to R.E. Beresneva, O.N. Sumenkova and O.A. Kosareva for the opportunity
to fruitfully work on this problem. 

     All authors express their gratitude to I.D. Novikov Jr. for helpful discussions and assistance in preparing the article.

\end{acknowledgements}

\bibliography{WH_silhouette_inner_CMB}

\end{document}